\documentclass[conference]{IEEEtran}
\ifCLASSINFOpdf

\else

\fi
\usepackage{subfig}
\usepackage{float}
\usepackage{graphicx}
\usepackage{algorithm}
\usepackage{algorithmicx}
\usepackage{algpseudocode}
\usepackage{setspace}
\usepackage{amsmath}

\begin{document}

\title{CFT: A Cluster-based File Transfer Scheme for Highway VANETs}

\author{\IEEEauthorblockN{Quyuan Luo$^{1}$, Changle Li$^{1*}$, Qiang Ye${^2}$, Tom H. Luan${^3}$, Lina Zhu$^{1}$ and Xiaolei Han$^{1}$}
\IEEEauthorblockA{$^{1}$State Key Laboratory of Integrated Services Networks, Xidian University, Xi'an, Shaanxi, 710071 China \\
${^2}$School of Mathematical and Computational Sciences, University of Prince Edward Island, Charlottetown, PE,  C1A 4P3 Canada\\
${^3}$School of Information Technology, Deakin University, Melbourne, VIC 3125, Australia\\
$^*$E-mail: clli@mail.xidian.edu.cn}
}
\maketitle

\begin{spacing}{1.0}
\begin{abstract}

Effective file transfer between vehicles is fundamental to many emerging vehicular infotainment applications in the highway Vehicular Ad Hoc Networks (VANETs), such as content distribution and social networking. However, due to fast mobility, the connection between vehicles tends to be short-lived and lossy, which makes intact file transfer extremely challenging. To tackle this problem, we presents a novel Cluster-based File Transfer (CFT) scheme for highway VANETs in this paper. With CFT, when a vehicle requests a file, the transmission capacity between the resource vehicle and the destination vehicle is evaluated. If the requested file can be successfully transferred over the direct Vehicular-to-Vehicular (V2V) connection, the file transfer will be completed by the resource and the destination themselves. Otherwise, a cluster will be formed to help the file transfer. As a fully-distributed scheme that relies on the collaboration of cluster members, CFT does not require any assistance from roadside units or access points. Our experimental results indicate that CFT outperforms the existing file transfer schemes for highway VANETs.

\end{abstract}







%



\IEEEpeerreviewmaketitle

\section{Introduction}

Vehicular Ad Hoc Networks (VANETs) have attracted much attention in both academia and industry. Among the problems to be tackled in VANETs, effective vehicle-to-vehicle (V2V) file transfer is fundamental to many emerging infotainment applications, such as content distribution and file sharing. However, it remains to be a very challenging problem because inter-vehicle links tend to be short-lived and lossy thanks to fast node mobility. With short-lived and lossy links, it is very likely that only part of a file, which often turns to be useless in the end, can be transferred. This leads to a significant waste of network resources and hurts user experience seriously.

To enable efficient file transfer in VANETs, a variety of different schemes have been proposed. Deng \emph{et al.} \cite{deng2016cooperative} proposed a Prior-Response-Incentive-Mechanism to stimulate vehicles to take part in cooperative downloading in VANETs-LTE heterogeneous networks. Huang \emph{et al.} \cite{huang2016ecds} proposed a cell-based clustering scheme and a strategy of inter-cluster Relay Selection to construct a pear-to-pear (P2P) network of scale-free property, which help to enhance the information spread. Ota \emph{et al.} \cite{ota2015mmcd} proposed a cooperative downloading algorithm called Max-throughput and Min-delay Cooperative Downloading, in which the RoadSide Units (RSUs) intelligently select vehicles to serve towards the minimal average delivery delay of file transfer. Yang \emph{et al.} \cite{yang2013maxcd} proposed a Cooperation-aided Max-Rate First method, in which the roadside unit selects the node with the highest data rate as the receiver to serve. Liu \emph{et al.} \cite{liu2012cooperative} proposed a cooperative downloading strategy that can provide mobile users with varied services to access the Internet using WiFi according to user-defined classes in highway scenarios. T. Wang \emph{et al.} \cite{wang2013dynamic} proposed a cooperative approach based on coalition formation games, in which on-board units download pieces from RSUs and then exchange their possessed pieces by broadcasting to and receiving from their neighbors.


Most of the existing file transfer schemes focus on the QoS issues, such as packet delay and network throughput. Despite the importance of the QoS issues, we believe that file transfer integrity (i.e. whether a file can be completely transferred or not) plays a more important role because it has a direct impact on the quality of user experience. Normally, users can tolerate some extra delay. However, if file transfer attempts often lead to failures, users will be seriously upset. So far, only few studies on file transfer in VANETs have paid enough attention to the integrity aspect.
Luan \emph{et al.} \cite{luan2013integrity} proposed an integrity-oriented content transmission scheme, which focuses on the performance of the entire file transfer process and is optimized towards guaranteed integrity. However, with the scheme presented in \cite{luan2013integrity}, the file that is unlikely to be completely transferred during the transient connection time will simply be discarded, which is overly simple and leaves much room for improvement.

In this paper, we present a high-integrity Cluster-based File Transfer (CFT) scheme for highway VANETs. CFT adopts a cooperative approach and achieves high-integrity file transfer between vehicles without the assistance of roadside units or access points. With CFT, when the requested file cannot be successfully transferred from the source to the destination over a single direct V2V connection, a cluster is formed and the file is collaboratively transmitted over multiple hops. To facilitate the multi-hop file transfer, we developed a \emph{Connection Time Prediction Model} and a \emph{Content File Distribution Model} to evaluate the transmission capability of cluster members and optimally select the intermediate relay nodes.



The rest of the paper is organized as follows. Section II presents the models adopted in CFT and Section III describes the details of the proposed cluster-based file transfer scheme. Section IV includes our experimental results and Section V concludes the paper with closing remarks.

\section{System Model}

In this paper, we focus on the scenario in which vehicles travel on a bi-directional highway. We assume that all vehicles are equipped with the Global Positioning System (GPS) and are aware of their geographical locations.

In our research, we adopted three models to study the file transfer problem in VANETs:  vehicle mobility model, connection time prediction model,  and vehicle-to-vehicle communication model \cite{zhou2014chaincluster}. The details of these models are presented as follows.

\subsection{Vehicle Mobility Model}
Considering the mobility features on practical highways, we apply the free mobility model to represent the mobility of vehicles on highways \cite{mao2013road}.


The mobility features of vehicles are summarized as follows:

\begin{itemize}

    \item The current moving speed is independent of the previous moving speed. In addition, the speed range is specified by a maximal speed and a minimal speed.

    \item A Safety Distance (SD) is defined. Namely, two adjacent vehicles on the same lane should maintain the SD for safety purposes. When the distance between two adjacent vehicles is less than SD, the rear vehicle should slow down until the distance between them meets at the SD requirement.

    \item A vehicle only travels along one lane of the highway.

\end{itemize}

\begin{figure}[!htb]
\fbox{
\begin{minipage}[t]{0.48\linewidth}
\centering
\includegraphics[width=1.7in]{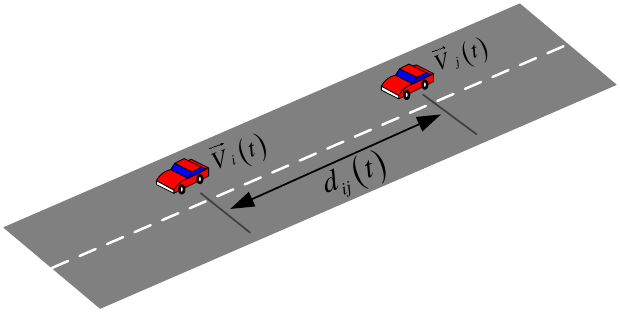}
\caption{Speed and Distance}
\label{fig mobility model}
\end{minipage}%
\begin{minipage}[t]{0.48\linewidth}
\centering
\includegraphics[width=1.3in]{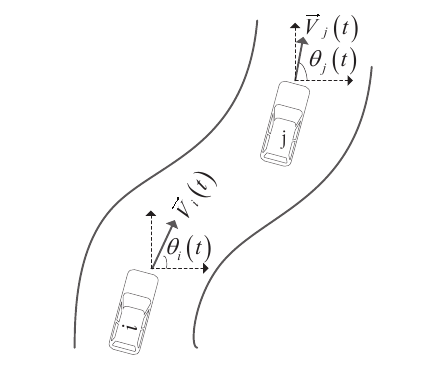}
\caption{Connection Time Prediction}
\label{fig prediction model for communication connection time}
\end{minipage}
}
\end{figure}

In the mobility model adopted in our research, both the speed of all vehicles and the distance between two adjacent vehicles are specified. Fig.~\ref{fig mobility model} illustrates the specifications for two adjacent vehicles.








Let $\Omega$ denote the set of vehicles. According to the mobility features mentioned previously, we can use the following velocity equations to represent the movement of two vehicles and the relationship between the speed of two vehicles:
\begin{equation}
\label{eq velocity exprission}
\small
\left\{ {\begin{array}{*{20}{l}}
{\left| {{{\vec V}_i}\left( {t + \Delta t} \right)} \right| = \left| {{{\vec V}_i}\left( t \right)} \right| + {\gamma _i}\left( t \right)*\left| {{{\vec a}_i}\left( t \right)} \right|*\Delta t},\\
{{V_{\min }} \le \left| {{{\vec V}_i}\left( t \right)} \right| \le {V_{\max }}},\\
{\left| {{{\vec V}_j}\left( t \right)} \right| \le \left| {{{\vec V}_i}\left( t \right)} \right|,{d_{ij}}\left( t \right) \le SD.{\rm{ }}}
\end{array}} \right.
\end{equation}
where ${{\vec V}_i}\left( {t} \right)$ represents the velocity vector of vehicle $i\in\Omega$ at time $t$, ${\gamma _i}\left( t \right)$ is a random number, ${{\vec a}_i}\left( t \right)$ is the acceleration vector of vehicle $i$ at time $t$, $\Delta t$ is the time interval, ${d_{ij}}\left( t \right)$ is the distance between vehicle $i$ and vehicle $j\in\Omega$ at time $t$, $SD$ denotes the safety distance between two vehicles.

With these velocity equations, we can arrive at a highway mobility model in terms of both time and space. In our research, we assume that the initial distance between any two vehicles is $d_{ij}(t_{0})$, and the initial speed of vehicles is set to $|{{\vec V}}({t_0})|$, and $\gamma$ is a random number between 0 and 1. Then we have the following equations:
\begin{equation}
\small
\left\{ \begin{array}{l}
{d_{ij}}({t_0}) = (1{\rm{ + }}\gamma ){\rm{*}}SD,\\
|{{\vec V}}({t_0})| = {V_{\min }} + \gamma *({V_{\max }} - {V_{\min }}).
\end{array} \right.
\end{equation}

\subsection{Connection Time Prediction Model}
We assume that the communication range of each node is $R$. The position, the velocity and the moving direction of vehicle $i\in\Omega$ at time $t$ are (${x_i}(t),{y_i}(t)$), ${\vec V_i}(t)$ and ${\theta _i}(t)$ respectively. Similarly, the position, the velocity and the moving direction of vehicle $j\in\Omega$ at time $t$ are (${x_j}(t),{y_j}(t)$), ${\vec V_j}(t)$ and ${\theta _j}(t)$ respectively. Fig. \ref{fig prediction model for communication connection time} illustrates how the connection time is predicted in our research.

In order to predict the connection time between two vehicles, we use the following equations to calculate the connection time $\Delta T_{ij}$ \cite{su2001mobility}:
\begin{equation}
\small
\left\{ \begin{array}{l}
\Delta {v_x} = \mid{{\vec V}_i}(t)\mid cos {\theta _i}(t) - \mid{{\vec V}_j}(t)\mid cos {\theta _j}(t),\\
\Delta {v_y} = \mid{{\vec V}_i}(t)\mid sin{\theta _i}(t) - \mid{{\vec V}_j}(t)\mid sin{\theta _j}(t),\\
\Delta {d_x} = {x_i}(t) - {x_j}(t),\\
\Delta {d_y} = {y_i}(t) - {y_j}(t),\\
{(\Delta {d_x} + \Delta {v_x}*\Delta T_{ij})^2} + {(\Delta {d_y} + \Delta {v_y}*\Delta T_{ij})^2} = {R^2},
\end{array} \right.
\end{equation}
\begin{equation}
\small
\left\{ \begin{array}{l}
A = \Delta {v_x}*\Delta {d_x} + \Delta {v_y}*\Delta {d_y},\\
B = \Delta {v_x}^2 + \Delta {v_y}^2,
\end{array} \right.
\end{equation}
\begin{equation}
\label{eq communication connection time}
\small
\Delta T_{ij} = \frac{{ - A + \sqrt {B*{R^2} - {{(\Delta {v_y}*\Delta {d_x} - \Delta {v_x}*\Delta {d_y})}^2}} }}{B}.
\end{equation}
\subsection{Vehicle-to-Vehicle Communication Model}
In this section, we analyze the transmission capacity of the physical layer in vehicle-to-vehicle communication \cite{parent2013semantic}. Considering the characteristics of DSRC communication \cite{kenney2011dedicated}, the received signal power obeys Rice Distribution initially, but with the increase of the distance between two vehicles, the received signal power obeys Rayleigh Distribution. In order to better simulate the envelope of fading signal in highway vehicular networking environment, we apply the Nakagami-m distribution \cite{cheng2007mobile}. The probability density function of the signal envelope $f(x;\mu,\Omega )$ is formulated as
\begin{align}  \small
 f(x;\mu,\Omega ) &= \frac{{2{\mu ^\mu }}}{{\Gamma (\mu ){\Omega ^\mu }}}{x^{2\mu  - 1}}\exp ( - \frac{\mu }{\Omega }{x^2}), \\  \Gamma (\mu ) &= \int_0^\infty  {{e^{ - x}}} {x^{\mu  - 1}}dx,
 \end{align}
where  $\mu$ is the signal fading index related to the surroundings and the distance between two communication vehicles \cite{Ta2009on}. In our research, we adopt the following reference values: $\mu$=0.74 when $d_{ij}\in[90.5,230.7]$; $\mu$=0.84 when $d_{ij}\in[230.7,588]$. $\Omega$ is the average received power before envelope detection. It can be defined as
\begin{equation}
\small
\Omega  = {P_t}{G_t}{G_r}\frac{{h_t^2h_r^2}}{{d_{ij}^\alpha L}},
\end{equation}
where $P_t$ is the transmission power, $G_t$ and $G_r$ are the transmission and reception antenna gain respectively, $h_t$ and $h_r$ are the  transmission and reception antenna length respectively, $L$ is the loss coefficient of the system, and $\alpha$ represents the path loss index \cite{mao2006wsn}.
Then we can calculate the probability density function of the signal to noise ratio (SNR) using the following equation:
\begin{equation}
\small
{P_r}(\frac{S}{{{N_r}}} \le x) = 1 - \frac{{\Gamma (\mu,\frac{\mu }{\Omega }{N_r}x)}}{{\Gamma (\mu )}},
\end{equation}
where $N_r$ is the thermal noise power, ${\Gamma (\mu,\frac{\mu }{\Omega }{N_r}x)}$ is formulated as
\begin{equation}
\small
{\Gamma (\mu,\frac{\mu }{\Omega }{N_r}x)}=\int_{\frac{\mu }{\Omega }{N_r}{x}}^\infty  {{e^{-x}}{x^{\mu  - 1}}dx}.
\end{equation}

Here, we assume that the transmitter of each node in vehicular environment supports $K$ modulation rates, $c_k$ is the $k^{th}$ modulation rate (${c_1} < {c_2} <  \cdots  < {c_k}$, $1\leq k\leq K$), and $v_k$ is the pre-set threshold. When ${v_k} \le \frac{S}{{{N_r}}} \le {v_{k + 1}}$, the module velocity is $c_k$. In addition, we set ${v_{K + 1}} = \infty$. Consequently, according to the equations mentioned previously, the transmission rate $c_k$ is selected with the probability
\begin{equation}
\label{eq communication capacity}
\small
\Pr \{ C = {c_k}\}  = \left\{ \begin{array}{l}
\begin{array}{*{20}{c}}
{\frac{1}{{\Gamma (\mu )}}({\Gamma _k} - {\Gamma _{k + 1}}),}&{1 \le k \le K - 1}
\end{array}\\
\begin{array}{*{20}{l}}
{\frac{{{\Gamma _k}}}{{\Gamma (\mu )}},}&{k = K},
\end{array}
\end{array} \right.
\end{equation}
\begin{equation}
\small
\label{eq communication capacity 0}
\Pr \{ C = 0\}  = 1{\rm{ - }}\sum\limits_1^K {\Pr \{ C = {c_k}\} },
\end{equation}
where ${\Gamma _k}$ and ${\Gamma _{k+1}}$ are defined as
\begin{equation}
\small
\left\{ \begin{array}{l}
{\Gamma _k}  \quad = \ \int_{\frac{\mu }{\Omega }{N_r}{v_k}}^\infty  {{y^{\mu  - 1}}{e^{-y}}dy}, \\
{\Gamma _{k + 1}} = \int_{\frac{\mu }{\Omega }{N_r}{v_{k + 1}}}^\infty  {{y^{\mu  - 1}}{e^{-y}}} dy.
\end{array} \right.
\end{equation}

Thus, the average transmission rate can be obtained through the following equation:
\begin{equation}
\small
\label{eq averange transmission rate}
\begin{aligned}
E(c) & = 0 \cdot {P_r}(C = 0) + \sum\limits_{i = 1}^K {{C_i} \cdot {P_r}(C = {C_i})} \\
 & = \sum\limits_{i = 1}^K {{C_i} \cdot {P_r}(C = {C_i})}.
\end{aligned}
\end{equation}



\section{CFT: A High-Integrity File Transfer Scheme}
In this section, we present the details of the proposed high-integrity file transfer scheme, CFT. Based on the established clusters, CFT leads to high file transfer integrity.
An overview of CFT is presented as follows:
\begin{itemize}
    \item When a vehicle needs a file, it will broadcast a resource request message to its neighboring vehicles. If a neighbor vehicle has the file, it will send a response message back. Thereafter, the request vehicle selects a neighboring vehicle as the downloading source according to the velocity, the direction of the vehicle and the communication capacity.
    \item If two vehicles can complete the file transfer within their connection time, the request vehicle downloads the file directly without establishing a cluster.
    \item If two vehicles cannot complete the file transfer within their connection time, the request vehicle establishes a linear cluster. Then the request vehicle cooperates with the vehicles in the cluster to download the file.
\end{itemize}

To evaluate the communication capacity between two vehicles and establish a cluster when necessary, each vehicle requires the position, the velocity and the moving direction its neighbors. Therefore, we assume that all vehicles in the VANETs under investigation are equipped with GPS. In addition, each vehicle broadcasts Hello messages periodically in order to collect the position and velocity information of its neighboring vehicles. The details of CFT are presented as follows.
\subsection{Transmission Capacity between Two Vehicles}
\begin{figure}[!htb]
\centering
\includegraphics[width=0.45\textwidth]{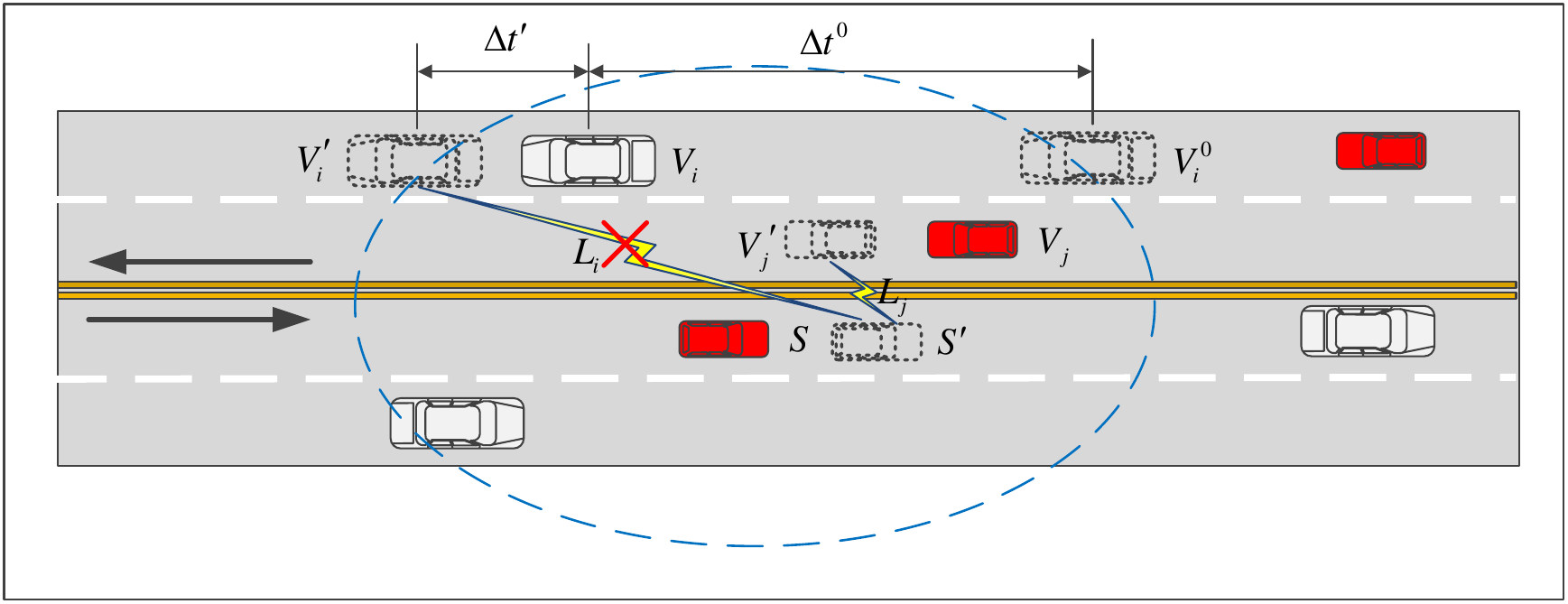}
\caption{Content File Distribution Model: $S$ selects another vehicle ($V_i$) to transfer fragments once the predicted amount of fragments have been transferred to the current cooperative vehicle ($V_j$).}
\label{fig_FileFragmentTransfer}
\end{figure}

In order to evaluate the transmission capacity between two vehicles, we developed the \emph{Content File Distribution Model} in our research. This model can be illustrated using Fig. \ref{fig_FileFragmentTransfer}.

If the request vehicle and resource vehicle cannot complete the file transfer within their connection time, cooperative vehicles are needed. The file content is equally divided into $N$ fragments denoted by $\Gamma=\{\gamma_1,\gamma_2,...,\gamma_N\}$ with the size of each fragment $s$. During the whole connection time $\Delta T_{i,S}$, vehicle $i\in \Omega$ can not exactly download $n$ fragments since vehicle $i$ is out of the communication range of the resource vehicle $S$ so that the link $L_{i}$ between $i$ and $S$ is disconnected when the $n^{th}$ fragment is transferring. And according the predicted connection time $\Delta T_{i,S}$, we can obtain the number of fragments $n_{i}$ by
\begin{equation}
\small
\label{eq n fragments}
{n_i} = \left\lfloor {\frac{{E(c) \cdot \Delta {T_{i,S}}}}{s}} \right\rfloor,
\end{equation}
where $\lfloor \cdot \rfloor$ denote the floor function, $E_(c)$ is the average transmission rate which can be obtained by Eq. \ref{eq averange transmission rate}. Besides, in Fig. \ref{fig_FileFragmentTransfer}, $\Delta {t^0}$ is the time spending on downloading $n_{i}$ fragments completely, $\Delta t'$ is the time $\Delta {T_{i,S}}$ minus $\Delta {t^0}$ and during which the $n^{th}$ fragment can not be downloaded completely. Their relationship is formulated as
\begin{equation}\label{eq shengyushijian t}
\small
\left\{ {\begin{array}{*{20}{l}}
{\Delta t' + \Delta {t^0} = \Delta {T_{i,S}}},\\
{\Delta {t^0} = \frac{{{n_i} \cdot s}}{{E(c)}}}.
\end{array}} \right.
\end{equation}
%
%

In our proposed scheme, upon transferring the $n_{i}^{th}$ fragment, $S$ selects another cooperative vehicle $j\in \Omega$ to transfer file fragments and establishes the link $L_{j}$. Through this method, such data loss $D_{loss}$ will not occur and it is of great importance for transfer time saving. Consequently, the communication capacity $C_{i,S}^{c}$ between any vehicle $i$ and resource vehicle $S$ is calculated as
\begin{equation}\label{eq communication capacity}
\small
 C_{i,S}^{c}=s \cdot \left\lfloor {\frac{{E(c) \cdot \Delta {T_{i,S}}}}{s}} \right\rfloor.
\end{equation}

\begin{figure}[!htb]
\centering
\includegraphics[width=0.45\textwidth]{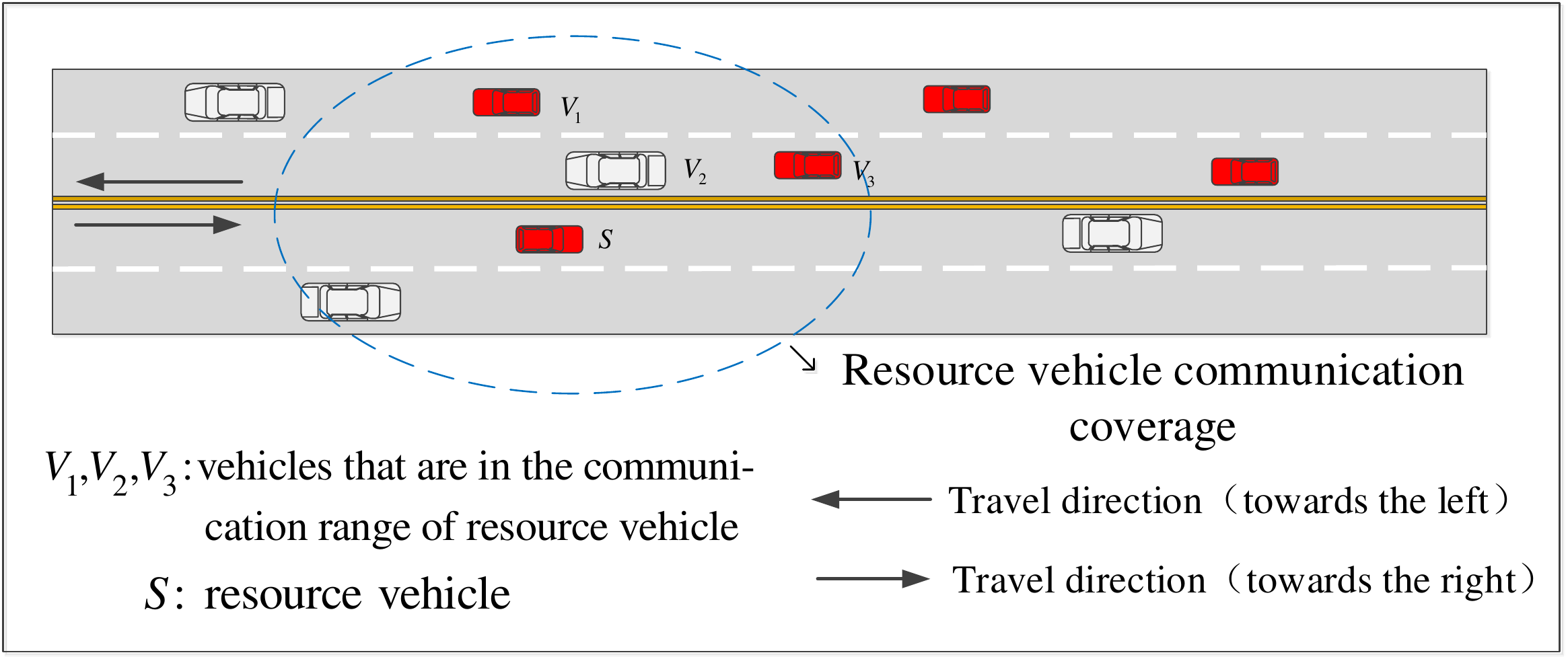}
\caption{Multiple Vehicles within Communication Range: the nearest vehicle
is selected as the cluster member, thereafter S transfers fragments to it.}
\label{fig Resource vehicle how to select cluster member}
\end{figure}

In the cooperative phase, if
several vehicles are in the communication range of resource
vehicle, as shown in Fig. \ref{fig Resource vehicle how to select cluster member}, the resource vehicle $S$ will transfer the file fragments to the vehicle with the shortest distance from $S$.
\vspace{-0.15in}

%
\subsection{Cluster Establishment}

With CFT, if a vehicle cannot download the required file completely from the resource vehicle within the connection time between them, then the vehicle will establish a liner cluster and cooperate with other vehicles in the cluster to download the file. There are many methods to establish a cluster in VANETs. The key problem is how to find the vehicles that have similar characteristics as cluster members. In the proposed scheme, the following steps are used to establish a cluster.

\emph{Step 1}: The request vehicle broadcasts a request packet for cooperative file transfer, then a neighboring vehicle which is within the communication range and willing to assist sends back an ACK. When the request vehicle receives the ACK, it will request the basic information, such as velocity and location, of the neighboring vehicle. Thereafter, the appropriate neighboring vehicle will be invited to join the cluster. Note that the neighboring vehicle is a cluster member while the request vehicle is the cluster-head.

\emph{Step 2}: The neighboring vehicle that joins the cluster continues to broadcast the request packet for cooperative file transfer and invites its neighbors to join the cluster. Then the basic information about the newly-added cluster member is forwarded to the cluster-head vehicle. Step 2 is repeated until enough cluster members have jointed the cluster.

According to the V2V communication model mention previously, we are able to calculate the amount of data every cluster member can download.  Assuming that the size of the  file to be transferred is $V_{file}$, the size of all file fragments that vehicle $i$ in the cluster can download is $V_{data}^i$ and the number of the required vehicles (i.e. the size of the cluster) is $N_c$, then $V_{data}^i$ and $N_c$ are formulated as
%
\begin{equation}
\small
V_{data}^i = C_{i,S}^{c},
\end{equation}
\begin{equation}
\small
N_{c} = \left\{ {\min \left\{ n \right\}|\sum\limits_{i = 1}^n {V_{data}^i \ge {V_{file}},n = 1,2, \cdots } } \right\}.
\end{equation}
%
\subsection{Cooperative Vehicles Transfer File Fragments to Request Vehicle}
After cluster members complete the file fragment transfer, they forward their fragments to the resource vehicle. We adopt the IEEE 802.11b DCF mechanism as the MAC protocol of the network and employ the RTS/CTS mechanism to avoid the hidden terminal problem. Furthermore, we calculate the back-off time using a constant back-off window size so that we can calculate the average transmission probability of each vehicle as
\begin{equation}
\small
\zeta  = \frac{2}{{W + 1}}.
\end{equation}

In order to calculate the success probability of packet transmission, we assume that $n$ nodes compete for one channel where $n$ obeys Poisson Distribution and its probability density function is formulated as
\begin{equation}
\small
{f_n}(x) = \frac{{{{(\rho {R_{cs}})}^x}}}{{x!}}\exp ( - \rho {R_{cs}}),
\end{equation}
where $\rho$ is a traffic density parameter, $R_{cs}$ is the diameter of carrier sense range of a vehicle. Then the probability that a node successfully sending packets in any slot can be calculated as
\begin{equation}
\small
{P_{suc}} = \frac{{n\zeta {{(1 - \zeta )}^{n - 1}}}}{{1 - {{(1 - \zeta )}^n}}}.
\end{equation}
Then the throughput of the MAC layer between two vehicles can be formulated as
\begin{equation}
\label{eq thoughputs}
\small
{R_{thr}} = \frac{{E[{V_{payload}}]}}{{E[length \ of\ a\ slot\ time]}} = {\frac{{{P_{suc}}{L_p}}}{T}}[1-(1-\zeta)^{n}],
\end{equation}
%
where $V_{payload}$ is the payload information transmitted in a slot time, $L_p$ is the average length of a packet, $T$ is the average length of a slot which is formulated in \cite{luan2013integrity}.

%

Consequently, the amount of data that can be transferred between cooperative vehicle $i$ and request vehicle $R$ within their connection time can be calculated using the connection time $\Delta T_{i,R}$, which can be obtained using Eq. \ref{eq communication connection time}, and the throughput $R_{thr}$, which can be obtained using  Eq. \ref{eq thoughputs}.
%
\subsection{Detailed File Transfer Process}
The detailed file transfer process is summarized as Alg. 1, where R is the request vehicle, S is the resource vehicle, $V_{data}^{RS}$ is the transmission capacity between R and S, and $V_{file}$ is the size of the file to be transferred.
\begin{algorithm}[!htb]
\renewcommand\arraystretch{1.1}
\footnotesize
\small \caption{\small CFT: Cluster-based File Transfer}
\begin{algorithmic}[1]
\If {R requests resource from S}
    \State R evaluates the transmission capacity $V_{data}^{RS}$;
    \If {$V_{data}^{RS}\geq V_{file}$}
        \State R downloads files directly;
    \Else
        \State R establishes a cluster and cooperates to download files with cluster members;
    \EndIf
\EndIf
\end{algorithmic}
\label{alg flow chart}
\end{algorithm}

\begin{figure}[!htb]
\centering
\subfloat[]{\includegraphics[width = 0.49\linewidth]{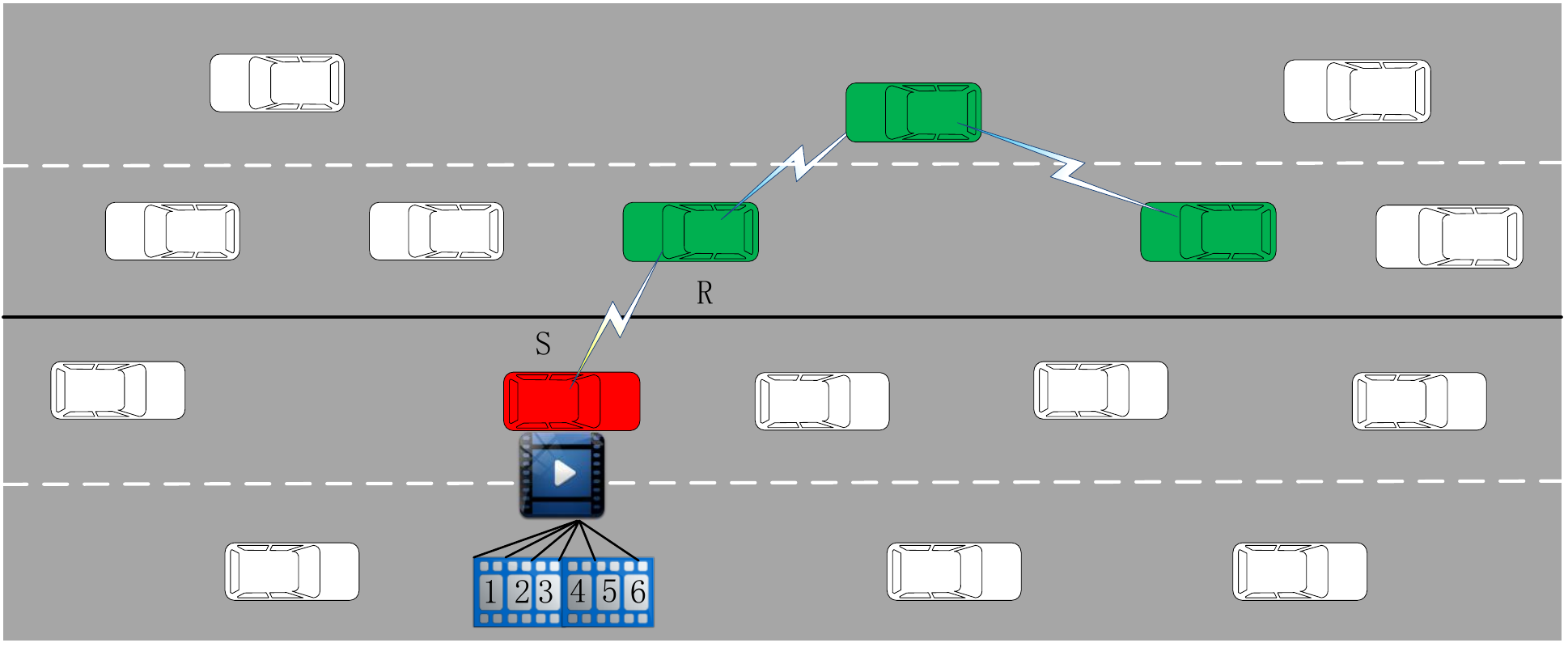}}\
\subfloat[]{\includegraphics[width = 0.49\linewidth]{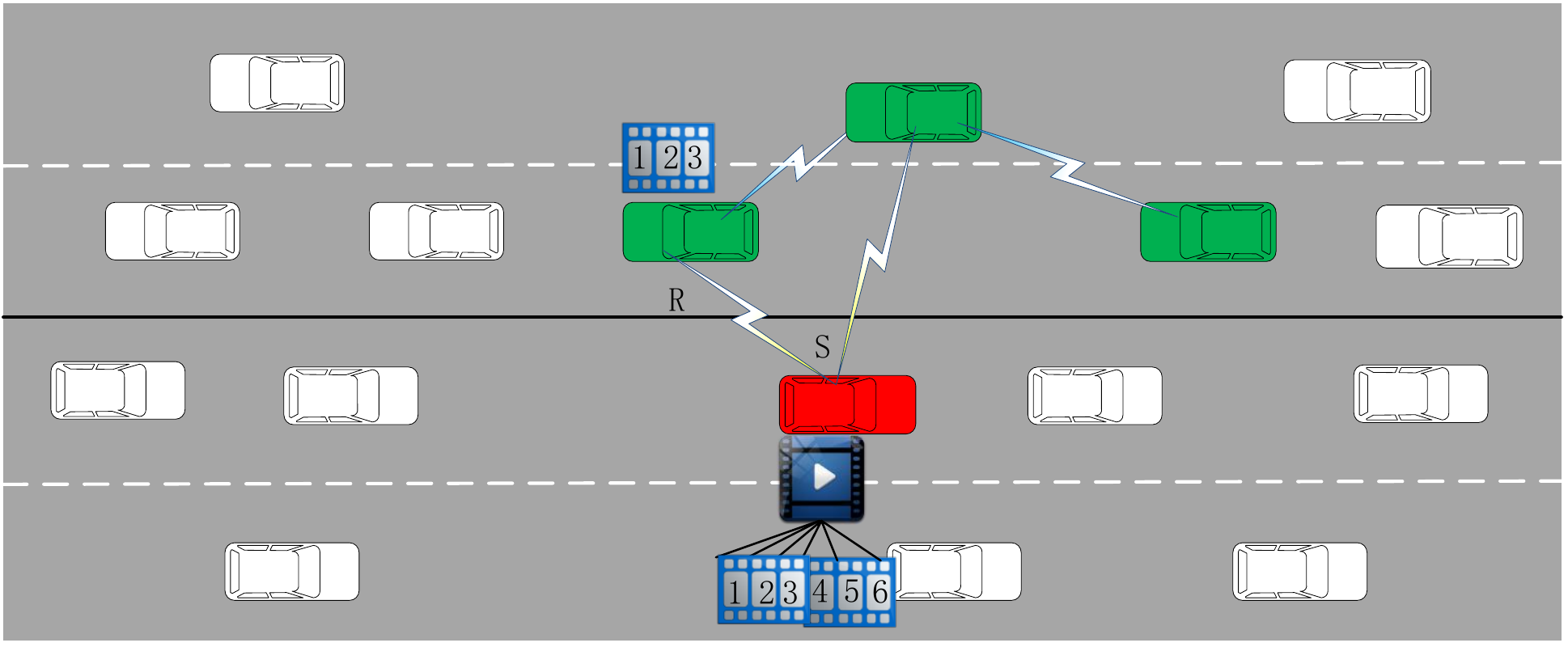}}\\
\subfloat[]{\includegraphics[width = 0.49\linewidth]{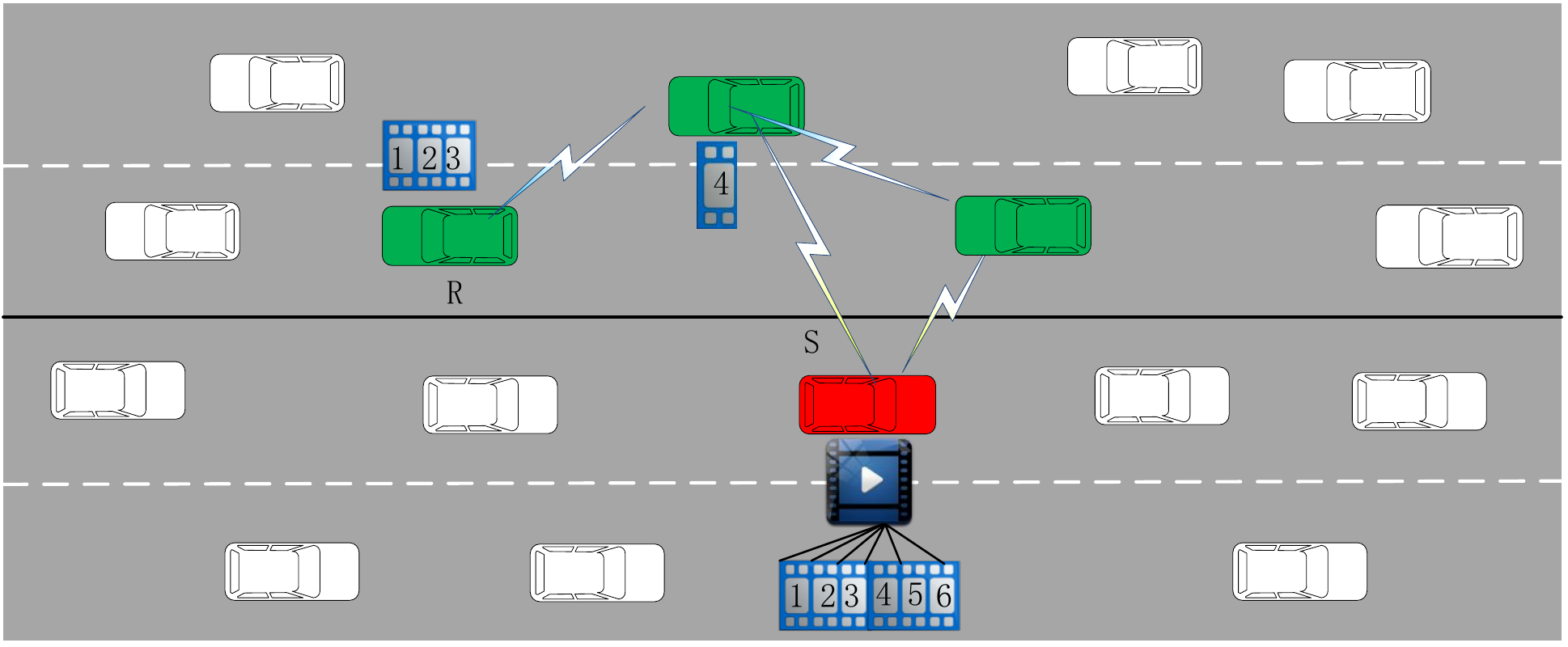}}\
\subfloat[]{\includegraphics[width = 0.49\linewidth]{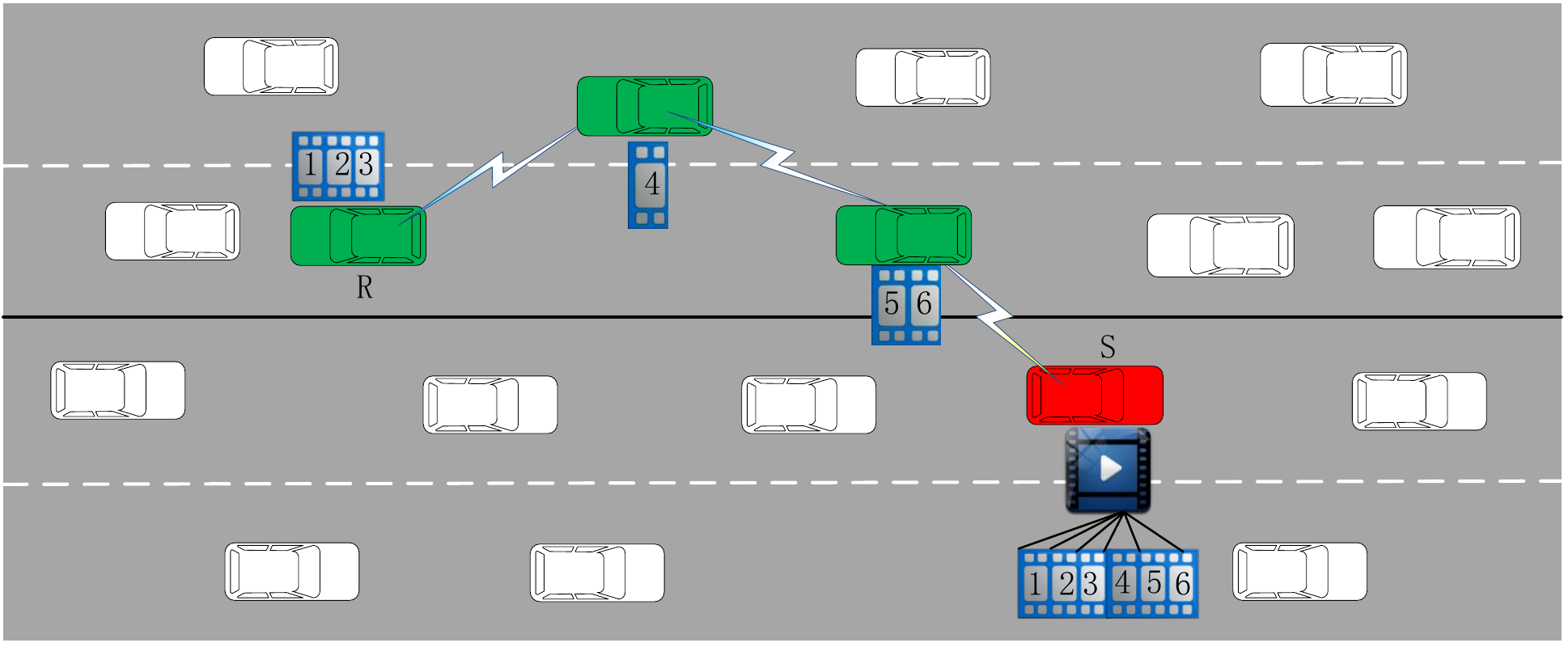}}\\
\caption{Cooperative Downloading}
\label{fig cooperative downloading}
\end{figure}

\begin{figure}[!htb]
\centering
\includegraphics[width=0.3\textwidth]{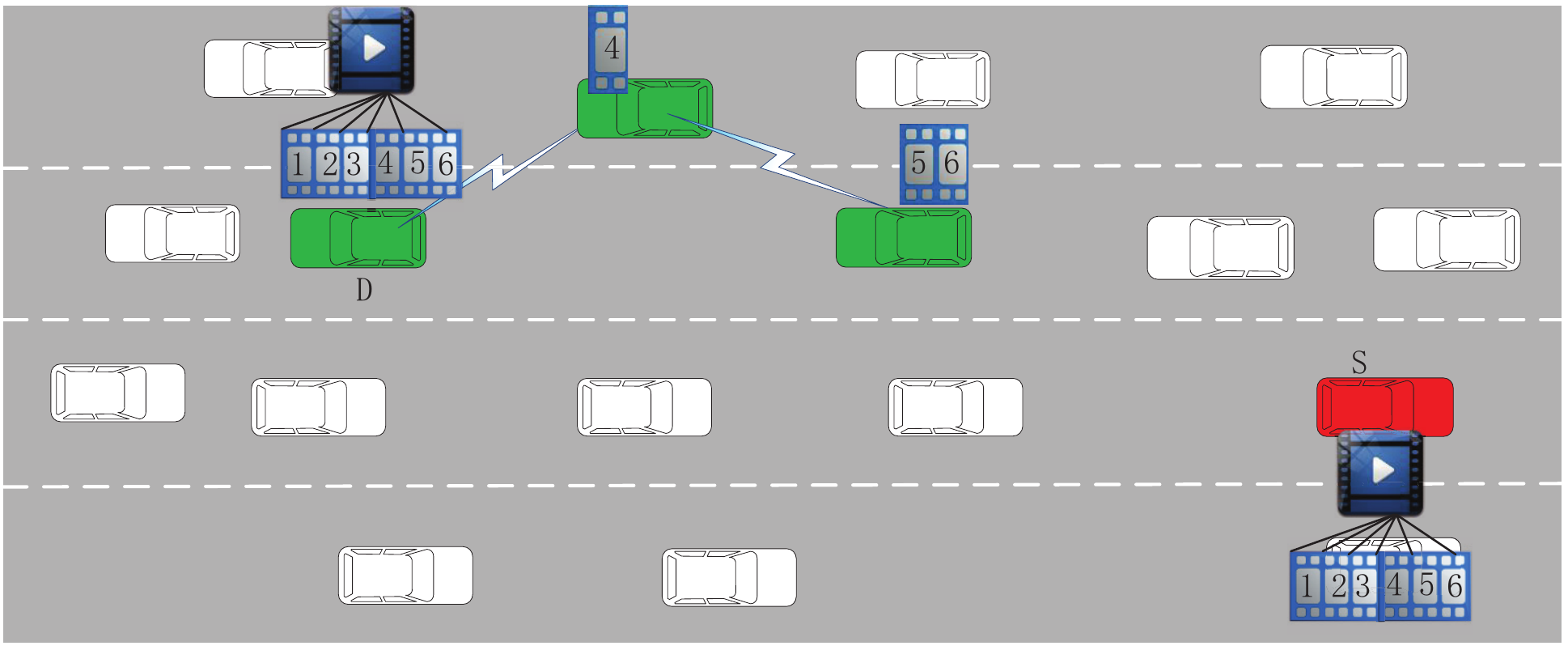}
\caption{Cluster Members Forward File Fragments to Cluster-Head}
\label{fig cluster members send files to the cluster-head R}
\end{figure}

Fig. \ref{fig cooperative downloading} includes a file transfer example. If R can download the file from S within the connection time, then R will not establish a cluster. Otherwise, R will establish a cluster for cooperative file transfer. We assume that three vehicles are needed to finish the file transfer in this example and each of them can finish the file fragment transfer within the connection time. After cluster members complete the file fragment transfer, they forward their fragments to the cluster-head R, which is illustrated in Fig. \ref{fig cluster members send files to the cluster-head R}.

%
\vspace{-0.2in}

\section{Experimental Results}
In our research, we studied the performance of CFT via extensive theoretical analysis and Mablab-based simulations. Our detailed experimental results are presented in this section. Specifically, the performance of CFT was investigated in terms of average connection time, average throughput, average transmission capability, maximum file transfer volume, and cluster size. We also compared CFT to one of the state-of-the-art scheme, IOCT \cite{luan2013integrity}, to understand the advantages and disadvantages of CFT. Our detailed experimental results are presented in this section.

\subsection{Simulation Settings}
In our simulations, we use a freeway model, where vehicles travel on a bi-directional highway with two lanes per direction \cite{mao2009graph}. The major  parameters are included in Table I. We use the IEEE 802.11b DCF mechanism as the MAC protocol and V2V communication protocol as the wireless communication protocol. In addition, the RTS/CTS mechanism is used to avoid the hidden terminal problem.

\begin{table}[!htb]
\renewcommand\arraystretch{1.1}
\footnotesize
\centering
\caption{Simulation Parameters}
\begin{tabular}{cc}
\hline
\textbf{Parameter}  & \textbf{Value}\\
\hline
Length of per lane ($km$) & $11$\\

Width of per lane ($m$) & $5$\\

Minimum speed ($km/h$) & $60$\\

Maximum speed ($km/h$) & $120$\\

Traffic density $\rho$ ($car/km$) & \{$5,6,\ldots,10$\}\\

Communication range of vehicle $R$ ($m$) & \{$250,300,\ldots,$600\}\\

Safety Distance $SD$ ($m$) & \{$75,100,150$\}\\

Size of back-off window $W$ & $32$\\

Length of A Packet $L_{p}$ ($KB$) & $4.2$\\

Length of a slot $T_{slot-time}$ ($us$) & $13$\\

Transmission time of RTS frame ($us$) & $53$\\

Transmission time of CTS frame ($us$) & $37$\\

Transmission time of DIFS frame ($us$) & $32$\\

Transmission time of SIFS frame ($us$) & $53$\\

Transmit power $P_t$ ($W$) & $0.2$\\

Hot noise power $N_r$ ($dBm$) & $-96$\\

Path loss index $\alpha$ & $4$\\

$G_t,G_r,h_t,h_r,L$ & $1$\\
\hline
\end{tabular}
\end{table}
\subsection{Simulation Results}
\begin{figure*}[!htb]

\begin{minipage}[t]{0.5\linewidth}
\centering
\includegraphics[width=2.2in]{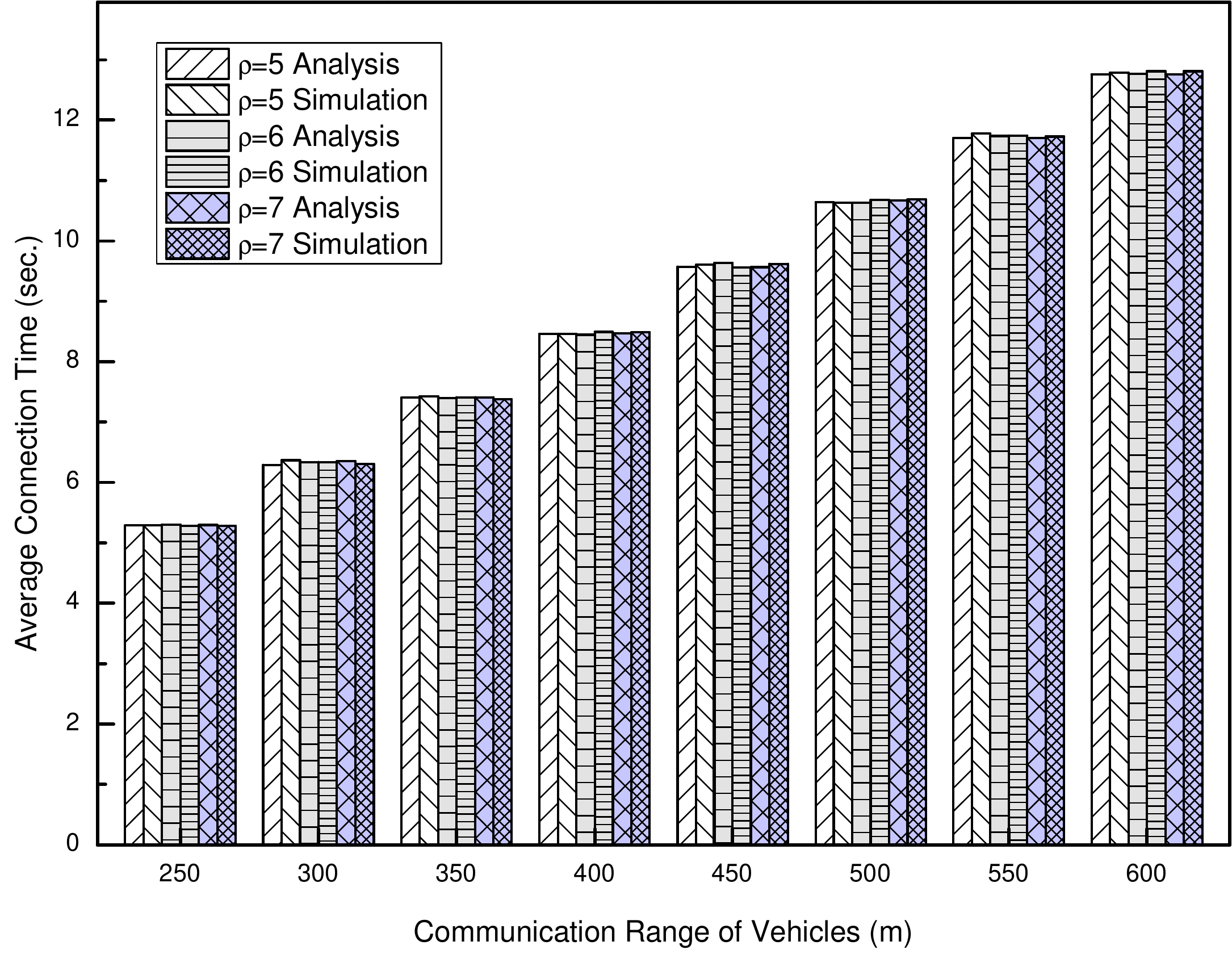}
\caption{Average Connection Time}
\label{fig Average communication connection time in different communication ranges}
\end{minipage}%
\begin{minipage}[t]{0.5\linewidth}
\centering
\includegraphics[width=2.2in]{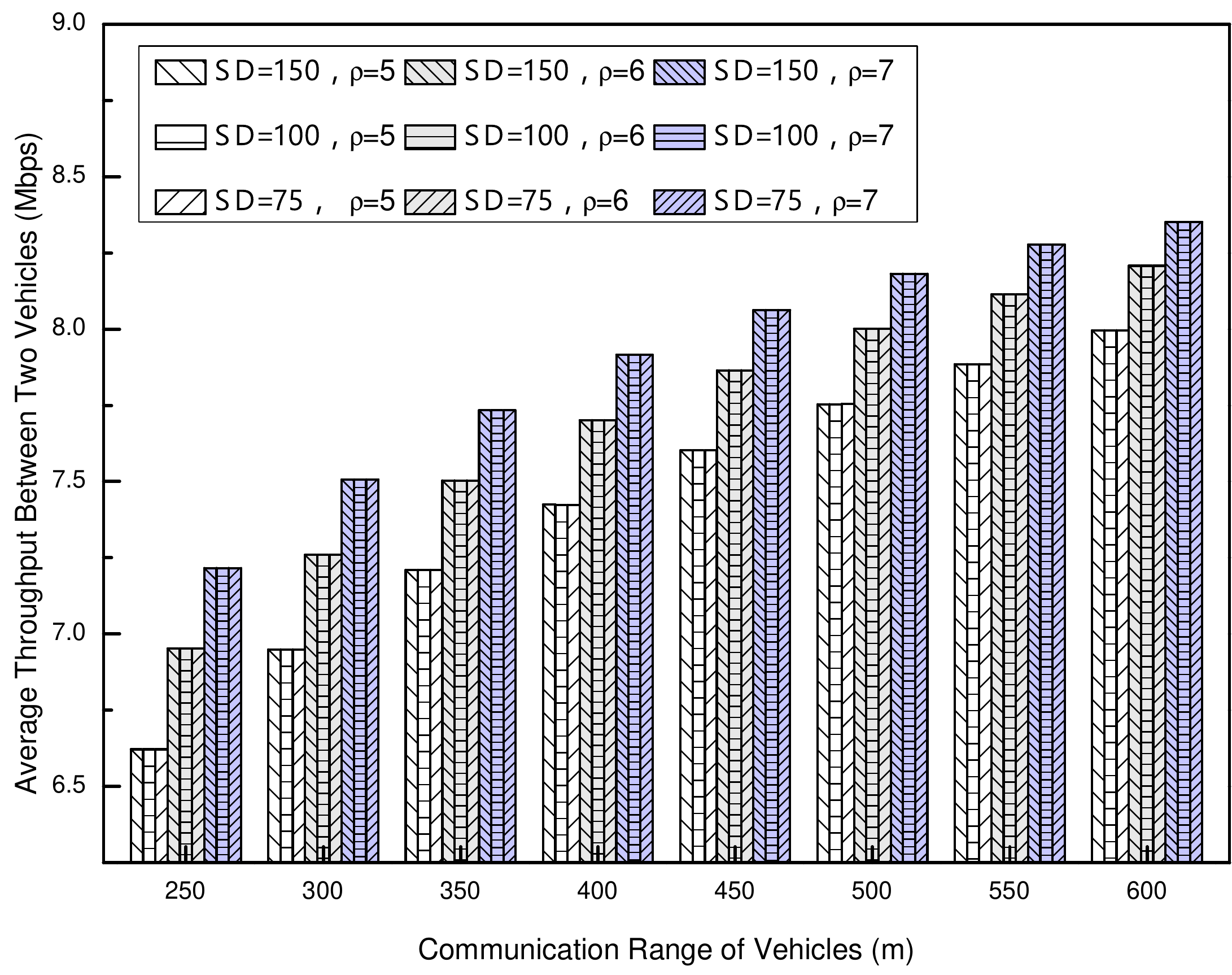}
\caption{Average Throughput}
\label{fig Average throughput in different  traffic densities and communication ranges}
\end{minipage}\\
\begin{minipage}[t]{0.33\linewidth}
\centering
\includegraphics[width=2.2in]{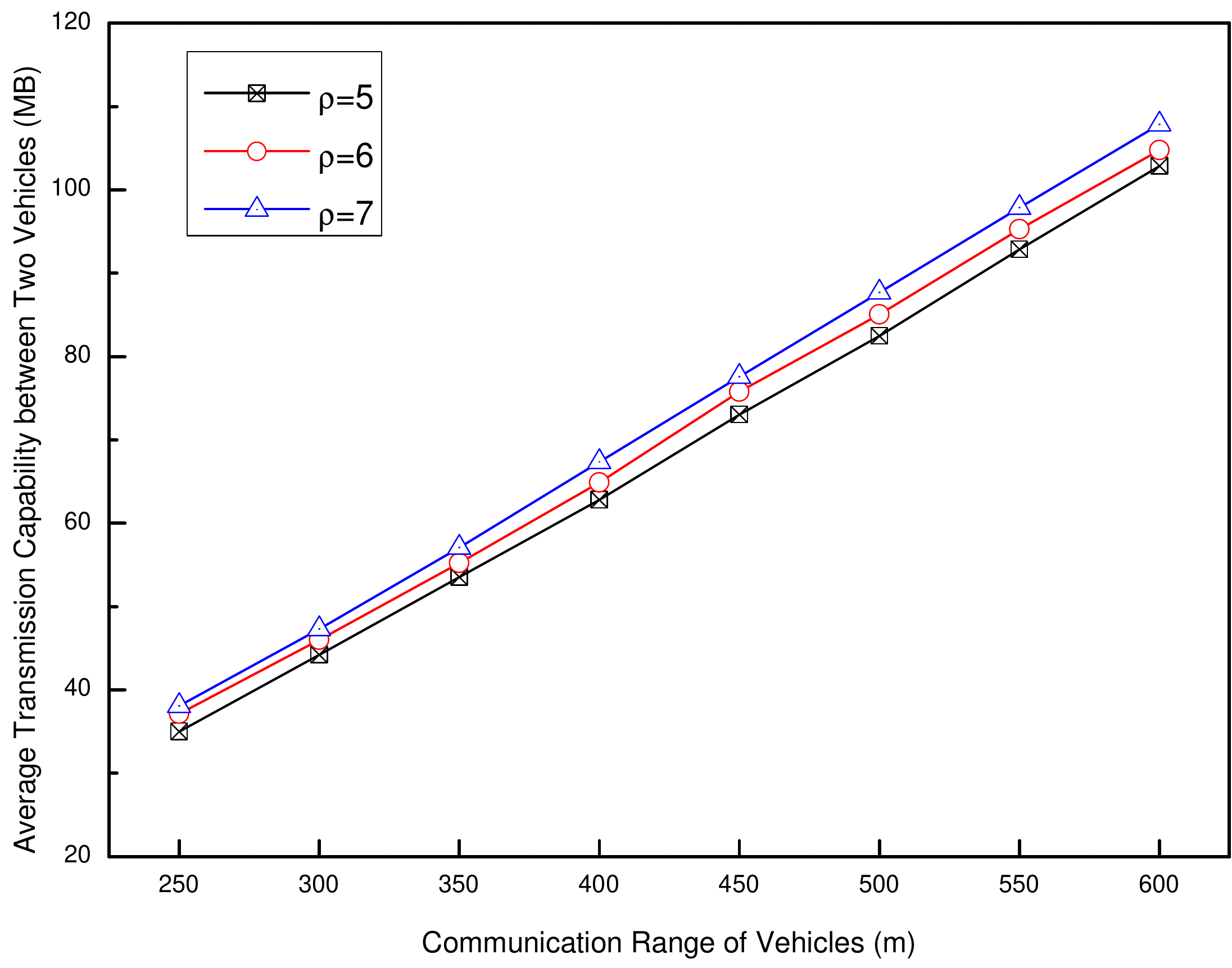}
\caption{Average Transmission Capability}
\label{fig Average transmission capability in different traffic densities and communication ranges}
\end{minipage}%
\begin{minipage}[t]{0.33\linewidth}
\centering
\includegraphics[width=2.2in]{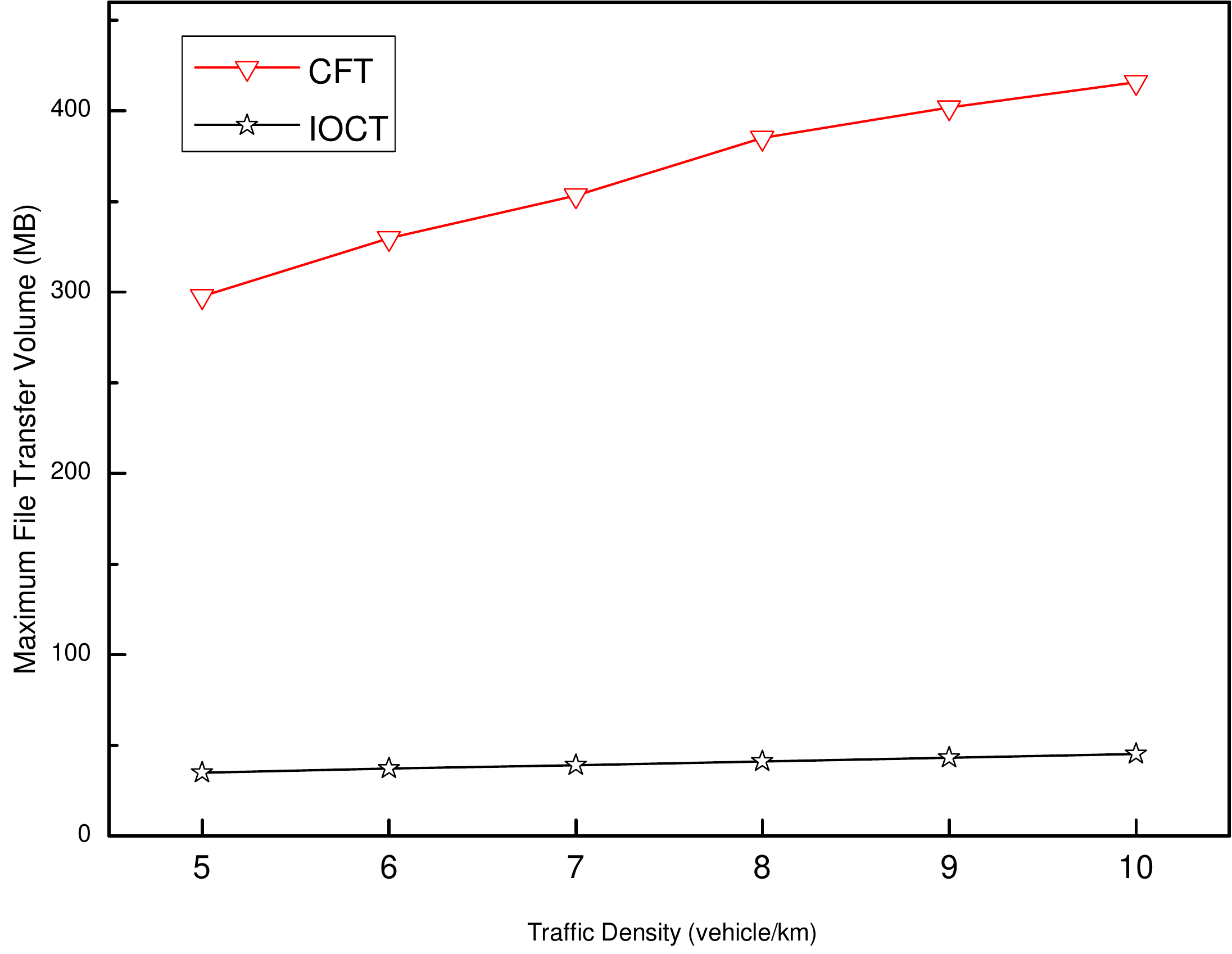}
\caption{Maximum File Transfer Volume}
\label{fig Cluster size and maximum file transfer volume in different file size}
\end{minipage}%
\begin{minipage}[t]{0.33\linewidth}
\centering
\includegraphics[width=2.2in]{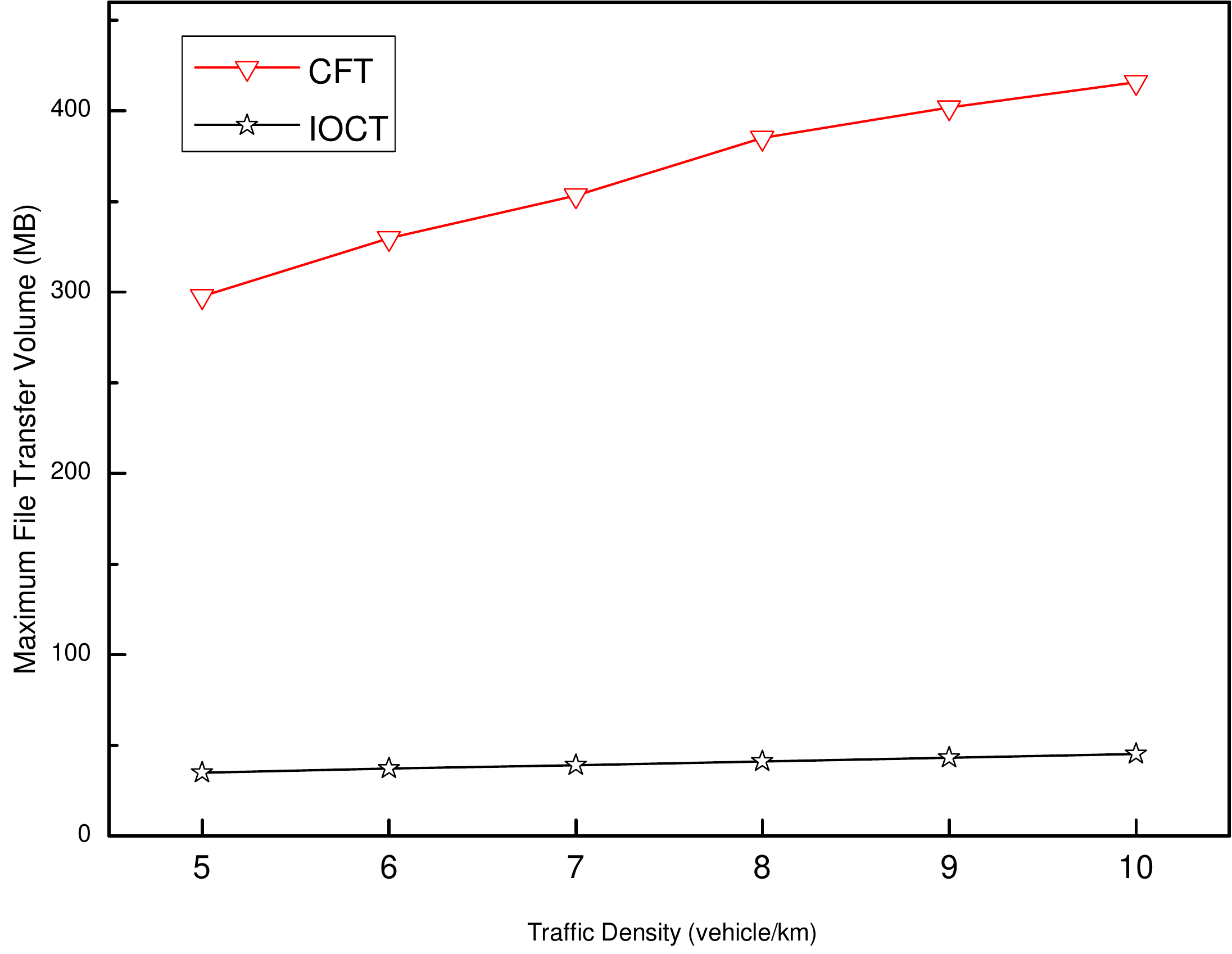}
\caption{Cluster Size}
\label{fig Cluster size in different file size}
\end{minipage}
\end{figure*}

%
\subsubsection{Average Connection Time}
Fig. \ref{fig Average communication connection time in different communication ranges} shows the average connection time under different traffic densities and communication ranges when $SD=150$ m. Note that $\rho$ denotes the number of vehicles per kilometer. Our analysis and simulation results indicate that the average connection time increases with the communication range. When the communication range is $250$ m, the average  connection time is 5.3 s. When the communication range is 600 m, the average connection time is about 12.7 s. The average connection time does not vary significantly with the densities under investigation.
%

\subsubsection{Average Throughput}
Fig. \ref{fig Average throughput in different traffic densities and communication ranges} shows the average throughput between two vehicles under different traffic densities, safety distances and communication ranges. Our experimental results indicate that when $\rho=5$ and the communication range varies from 250 m to 600 m, the average throughput varies from 6.6 Mbps to 8.0 Mbps; when $\rho=6$, the average throughput varies form 6.9 Mbps to 8.2 Mbps; when $\rho=7$, the average throughput varies form 7.2 Mbps to 8.4 Mbps. In summary, the average throughput increases with the traffic density and communication range.

\subsubsection{Average Transmission Capability}
Fig. \ref{fig Average transmission capability in different traffic densities and communication ranges} shows the average transmission capability between two vehicles under different communication ranges and traffic densities when $SD=150$ m. Our experimental results indicate that when $\rho=5$ and the communication range varies from 250m to 600 m, the average transmission capability varies from 35.0 MB to 102.9MB; when $\rho=6$, the average transmission capability varies from 37.1 MB to 104.8 MB; when $\rho=7$, the average transmission capability varies from 38.1 MB to 107.9 MB Therefore, the average transmission capability increases with the traffic density and linearly increases with the communication range.
%
%
\subsubsection{Maximum File Transfer Volume}
Fig. \ref{fig Cluster size and maximum file transfer volume in different file size} shows the maximum file transfer volume of CFT and IOCT under different traffic densities  when $R=250m$. Our experimental results indicate that when $\rho$ varies from $5$ to $10$, the maximum file transfer volume of IOCT varies from 35 MB to 45 MB, the maximum file transfer volume of CFT varies from 295 MB to 415 MB. The reason why the maximum file transfer volume of CFT is much greater is that a file can be transferred through multiple cluster members. What¡¯s more, we can see from Fig. \ref{fig Cluster size and maximum file transfer volume in different file size} that the maximum file transfer volume of IOCT is not sensitive to traffic density, which is because IOCT only
involves two vehicles.

\subsubsection{Cluster Size}
Fig. \ref{fig Cluster size in different file size} shows the average cluster size under different file sizes. Our experimental results indicate that the average cluster size increases with the file size. When $\rho=10$, the average cluster size varies from 1.7 to 13.2. Given the same file size, lower traffic density leads to a greater required cluster size.

%


%
\section{Conclusion}
In this paper, we present a high-integrity file transfer scheme for highway VANETs, CFT, which is based on the V2V communication model and  connection time prediction model. With CFT, if the request vehicle can download a file completely within the connection time between the request vehicle and the resource vehicle, it will download the file directly. If not, the request vehicle will establish a liner cluster, where the cluster members cooperate with each other to download the file. Through extensive simulations, we studied the performance of CFT in terms of average connection time, average throughput, average transmission capability, maximum file transfer volume, and cluster size. Our simulation results indicate that CFT outperforms the state-of-the-art file transfer schemes for highway VANETs.




%
\end{spacing}
\end{document}